\documentclass[doublecol]{epl2}
\usepackage{amsmath, amssymb, graphicx}

\newcommand{\eq}[1]{eq.~(\ref{#1})}
\newcommand{\fig}[1]{fig.~\ref{#1}}

\newcommand\aps{Adv. Polym. Sci.}
\newcommand\el{Europhys. Lett.}
\newcommand\pra{Phys. Rev. A}
\newcommand\pre{Phys. Rev. E}
\newcommand\prl{Phys. Rev. Lett.}
\newcommand\jcp{J. Chem. Phys.}
\newcommand\molp{Mol. Phys.}
\newcommand\rpp{Rep. Prog. Phys.}
\newcommand\mac{Macromol.}
\newcommand\lang{Langmuir}
\newcommand\sci{Science}

\title{Reversible gelation and dynamical arrest of dipolar colloids}
\shorttitle{Gelation of dipolar colloids}

\author{R. Blaak, M.~A. Miller and J.-P. Hansen}
\shortauthor{R. Blaak \etal}
\institute{University Chemical Laboratory, Lensfield Road, Cambridge CB2 1EW, United Kingdom}


\abstract{
We use molecular dynamics simulations of a simple model to show that dispersions of slightly elongated colloidal particles with long-range dipolar interactions, like ferrofluids, can form a physical (reversible) gel at low volume fractions.  On cooling, the particles first self-assemble into a transient percolating network of cross-linked chains, which, at much lower temperatures, then undergoes a kinetic transition to a dynamically arrested state with broken ergodicity.  This transition from a transient to a frozen gel is characterised by dynamical signatures reminiscent of jamming in much denser dispersions.
}

\begin{document}

\maketitle

Colloidal dispersions have been observed to undergo a glass transition to a disordered solid upon increasing their volume fraction $\phi$ \cite{Pusey87a}, or to undergo clustering and then gelation at low $\phi$ upon increasing the mutual attraction between particles \cite{Segre01a}.  Reversible gelation of colloidal dispersions is characterised by long-lived physical (rather than irreversible chemical) bonding between particles \cite{DelGado04}, which leads to the formation of sample-spanning networks capable of sustaining weak stresses at low volume fractions. In recent experimental and theoretical investigations the gelation process is almost invariably associated with short-range attractive interactions between particles, induced, e.g., by the polymer-driven depletion mechanism, which lead to low coordination clustering or jamming of preformed clusters \cite{Segre01a, Campbell05a, Manley05a, DelGado05a, Zaccarelli06a}.

In this Letter we explore an alternative gelation mechanism driven by long-range aniso\-tropic interactions resulting from electric or magnetic dipoles as in the technologically important ferrofluids.  Spherical particles with a point dipole are known to form a ``string phase'' at low volume fraction, comprising long transient chains of particles concatenated in energetically favourable head-to-tail configurations \cite{Weis05b}.  Individual chains percolate in such a phase, i.e., the average chain length diverges and in simulations the particles become connected to their own periodic images through the boundary conditions on the simulation cell.  However, the chains do not appear to be sufficiently interconnected to form a genuine gel-like network. Here, we show that network formation may be achieved at low $\phi$ by a modest elongation of the particles and their dipoles to make short dipolar dumbbells.

Chain formation in dipolar colloids is a natural consequence of the anisotropy of the long-ranged dipole--dipole interactions.  Colloidal chaining can also be achieved using short-ranged attraction, by limiting each particle to a maximum of two (reversible) bonds at specified sticky patches on their surface.  To form networks of such particles, it is necessary to introduce a second type of particle with higher valency to act as chain junctions \cite{Zaccarelli06a}.  Alternatively, a low coordination number can be enforced by the introduction of three-body forces that discourage small angles between the bonds to any given particle \cite{DelGado05a}.  In contrast, junctions arise with decreasing temperature in our one-component dipolar dumbbell fluid simply because their statistical weight increases relative to that of disconnected chains.

The nature \cite{Tlusty00a} and location of any first-order fluid--fluid phase transition in dipolar spheres is still not completely resolved.  The critical point of the Stockmayer fluid (Lennard-Jones particles with a point dipole) has been shown to drop rapidly in temperature as the isotropic attraction is switched off, leaving soft dipolar spheres \cite{Leeuwen93a}.  In contrast, Monte Carlo simulations of dipolar hard spheres indicate as many as three disordered phases at low temperature \cite{Camp00a}.  Less controversially, it has been shown that hard spherocylinders \cite{McGrother96b} or dumbbells \cite{Shelley99a} with an axial point dipole exhibit a gas--liquid transition within a limited range of particle elongations.  However, the present simulations of soft-core dumbbells with an extended dipole showed no sign of such a transition within the range of temperatures studied, as evidenced by the small wavevector behaviour of the static structure factor discussed later in the Letter.

\begin{figure}[tb]
\centerline{\includegraphics[width=85mm]{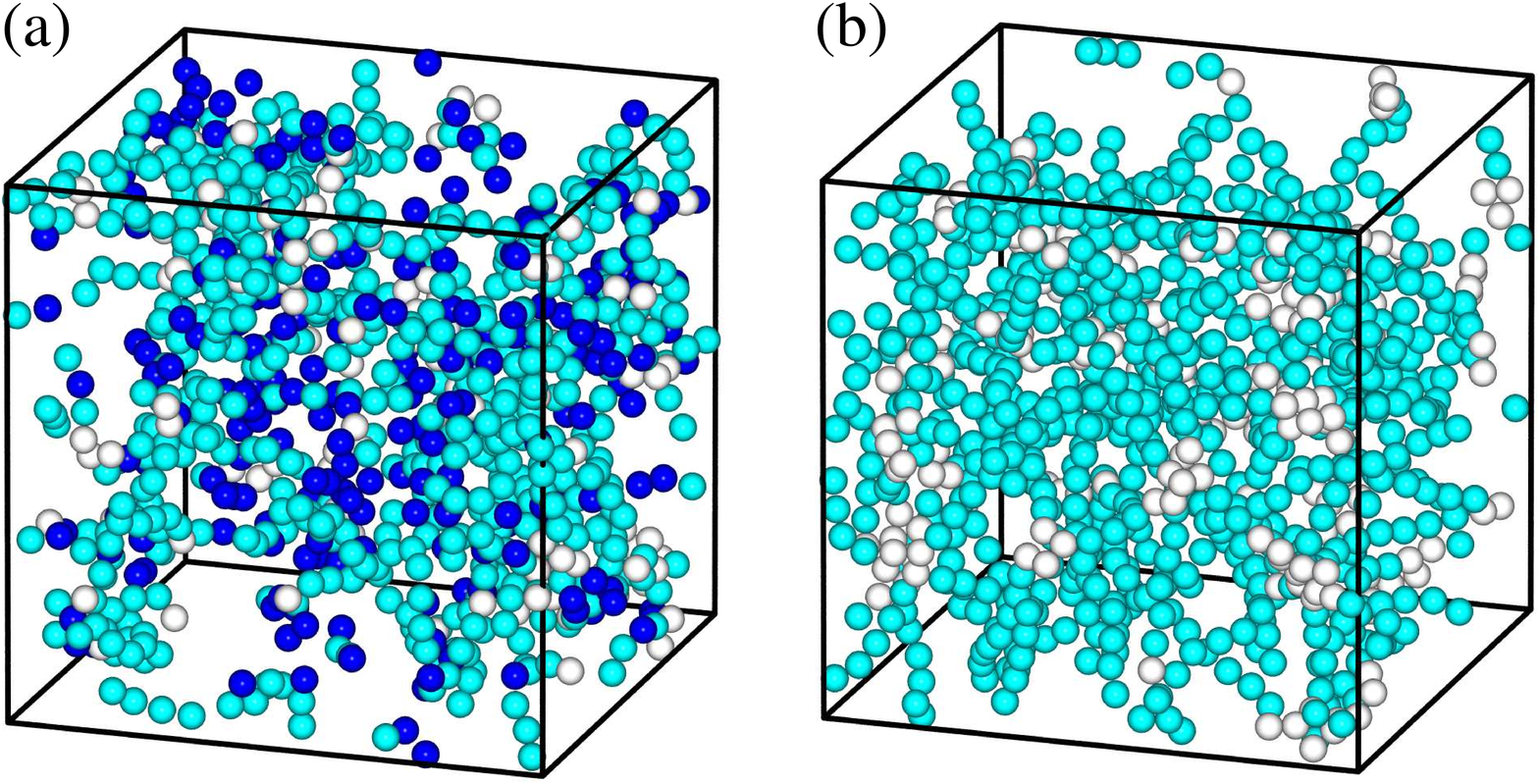}}
\caption{Simulation snapshots at (a) $T^* = 0.205$ and (b) $T^* = 0.041$. Dark, intermediate and light shading indicate particles with, respectively, one, two, and three or more neighbours. For simplicity, the slightly elongated dumbbells are represented by spheres.}
\label{F:snapshot}
\end{figure}

Our dumbbells are made up of two interpenetrating colloidal soft spheres carrying opposite charges $\pm q$ at their centres, separated by a fixed distance $d$. The resulting dipole moment is $\mu=qd$ and the point dipole limit is recovered as $d\to0$ and $q\to\infty$ at fixed $\mu$.  The interaction potential between sites on different dumbbells is the sum of a soft-sphere repulsion $c/r^{12}$, and a Coulomb interaction $\pm q^2/4 \pi\epsilon_0 \epsilon r$, where $r$ is the site--site distance, while $\epsilon_0$ and $\epsilon$ are the permittivity of free space and the relative permittivity of the medium, respectively. The finite extension of the dipole enhances string formation \cite{Ballenegger:2004MP}, while the dumbbell geometry facilitates branching of chains, which can thus interconnect into a genuine three-dimensional space-spanning network when the temperature $T$ is lowered at fixed $\phi$. A convenient length scale is the distance $\sigma$ between the centres of two dumbbells in the head-to-tail configuration of lowest total potential energy, while the natural energy scale is the dipole--dipole energy $u = \mu^2/4 \pi \epsilon_0 \epsilon \sigma^3$.  In terms of these quantities, the soft-core repulsion coefficient is fixed at $c=0.0208u\sigma^{12}$.  We shall use the reduced units $T^* = k_B T/u$ and $\rho^* = \rho \sigma^3$ ($\phi = \pi \rho^*/6$), where $k_B$ is Boltzmann's constant and $\rho = N/V$ is the number of particles per unit volume, while the reduced time is $t^* = t/t_0$ with $t_0 = \sqrt{m \sigma^2/u}$ and $m$ the particle mass. For typical colloidal particles in water, $\sigma \simeq 10^2$ nm, $m \simeq 10^{-18}$ kg, and $q \simeq 10^2$ proton charges so that for $d/\sigma=0.217$, as used in the present simulations, $u/k_B \simeq 10^5\,{\rm K}$ and $t_0 \simeq 10^{-7}\,{\rm s}$, while room temperature corresponds to $T^* \simeq 0.01$.

The constant-temperature molecular dynamics (MD) simulations were carried out on periodic samples of $N=1000$ dipolar dumbbells, using the Berendsen thermostat \cite{Berendsen:1972} and the standard Verlet leap-frog algorithm \cite{Book:Frenkel-Smit}, implemented in the Gromacs package \cite{Gromacs}.  Long-range Coulomb interactions are dealt with by the particle-mesh Ewald method.  The standard time step was chosen to be $\Delta t^* = 0.0008$ and most simulation runs extended over roughly $10^6$ time steps after initial equilibration, with a few runs of up to ten times this length at the lowest temperatures.  The Newtonian dynamics used here neglect the stochastic (Brownian) and hydrodynamic forces induced by the solvent.  No single technique without explicit solvent fully captures these effects, and Newtonian dynamics remains appealing because of its efficiency and lack of ambiguity.  It has also been shown, at least for denser systems, that long-time dynamical evolutions are not affected by the precise short-time dynamics \cite{Lowen:1991PRA,Gleim:1998PRL}.

\begin{figure}[tb]
\centerline{\includegraphics[width=80mm]{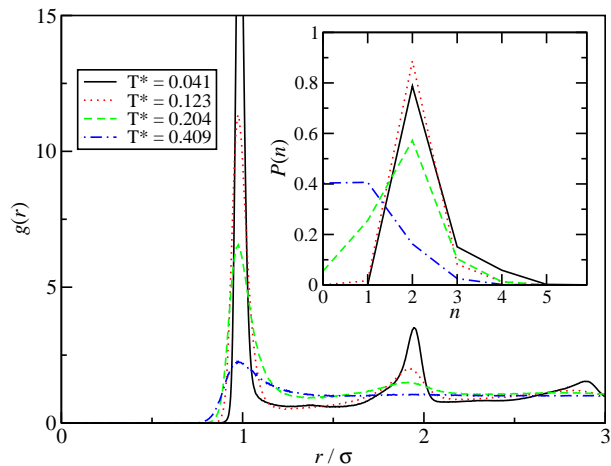}}
\caption{Radial distribution functions, $g(r)$ at four reduced temperatures $T^*$. Inset: probability distribution $P(n)$ of the number of neighbours $n$ per particle.  Lines join points at integer values of $n$ for clarity.}
\label{F:rdf}
\end{figure}

All simulations were carried out along the isochore $\phi = 0.0745$ and $T^*$ was gradually lowered from $T^* \simeq 0.4$ down to $T^* \simeq 0.02$. Typical snapshots of MD-generated configurations are shown in \fig{F:snapshot} for $T^*=0.205$ and $T^*=0.041$. The snapshots highlight singly-connected (end-of-chain), doubly-connected (mid-chain), and triply- or more highly-connected (junction) particles.  We determine the connectivity of two given particles by the simple geometric criterion that a pair of opposite charges---one on each dumbbell---lie closer than $r_{\rm cut}=\sigma$.  The higher temperature snapshot shows long chains which are hardly connected, while the lower temperature configuration is typical of a percolating network. The percolation threshold, where 50\% of configurations contain a system-spanning cluster, is found to occur at $T_p^*=0.177$.  At this temperature, the network is highly transient and the chains are only loosely connected.  Slightly below $T_p$, the fraction of percolating configurations rapidly rises to unity as the largest cluster becomes more ramified.  Nevertheless, the lifetime of individual connections between chains exceeds typical simulation time scales (i.e., millions of time steps) only at much lower temperatures.

The temperature dependence of the pair distribution function $g(r)$ of the particle centres is illustrated in \fig{F:rdf}.  At the highest temperature, a single peak appears at $r \simeq \sigma$, signalling the formation of dimers. As $T^*$ drops, the amplitude of the main peak rises sharply and well-separated secondary peaks develop at roughly multiple values of $\sigma$ as would be the case for a one-dimensional system, pointing to chain formation. The overall form of $g(r)$ differs considerably from that of radial distribution functions for dense, highly-coordinated fluids. The inset shows the probability distribution of the number $n$ of neighbours per particle. Above $T_p^*$ most particles have a single or no neighbour, while below $T_p^*$ almost all particles are doubly coordinated and a significant fraction of particles have three neighbours, indicating the existence of branched chains, which are crucial for network formation.

The Fourier transform of $g(r)$ yields the structure factor $S(q)$ which is experimentally accessible by X-ray or small-angle neutron diffraction. Except at the highest temperatures $S(q)$ exhibits a pronounced low-$q$ peak, indicative of clustering, similar to that observed for other models of gelation \cite{DelGado05a}.  Extrapolation of the MD data to $q=0$ provides an estimate of the isothermal compressibility $\chi_T$ via the standard fluctuation formula \cite{Book:Hansen-McDonald3}. The amplitude of the central peak implies that the gel is significantly more compressible than the corresponding gas of non-interacting particles. $S(q=0)$ appears to grow slowly as $T$ drops, but remains finite, precluding the possibility of spinodal instability associated with phase separation. The pressure is readily calculated from the virial theorem and turns out to drop rapidly with $T$ due to clustering, as expected.

Another important static characteristic is the relative dielectric permittivity $\epsilon$, which may be estimated from the fluctuations of the total dipole moment $\vect{M}$ of the sample using Kirkwood's relation \cite{Book:Hansen-McDonald3} adapted to metallic boundary conditions:
\begin{equation}
\epsilon=1+\frac{4\pi}{3}\frac{\rho\mu^2}{k_B T}\left\langle\frac{|\vect{M}|^2}{N\mu^2}\right\rangle.
\label{E:Kirkwood}
\end{equation}
$\epsilon$ exhibits a striking non-monotonic behaviour as a function of temperature. It first increases as $T$ drops, due to the $1/T$ factor in \eq{E:Kirkwood} and to enhanced dipolar fluctuations associated with cluster formation.  $\epsilon$ then reaches a maximum $\epsilon \simeq 10$ at $T^* \simeq 0.2$, before dropping sharply to values close to the ideal gas value $\epsilon=1$ at the lowest temperatures. Such low values show that dipolar fluctuations become strongly suppressed at very low temperatures, indicating incipient dynamical arrest.  At such temperatures, the topology of connections between the colloidal chains ``freezes'' into a quasi-permanent arrangement, thereby restricting the motion of particles to the scope permitted by deformations of the chains.

\begin{figure}[tb]
\centerline{\includegraphics[width=80mm]{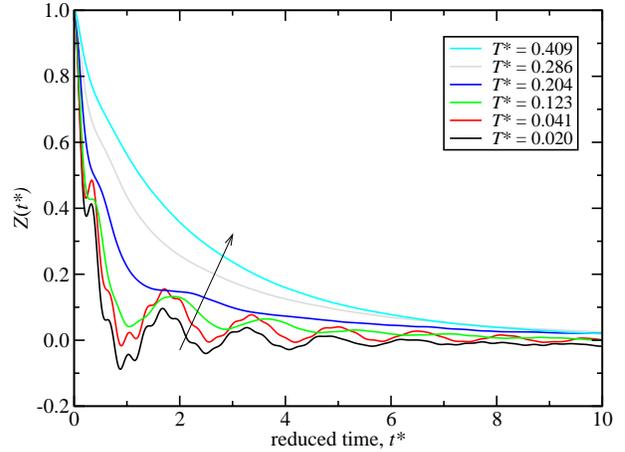}}
\caption{Velocity autocorrelation function $Z(t^*)$.  The arrow intersects the curves in order of increasing temperature.}
\label{F:Z}
\end{figure}

The expected dynamical slowing down at low temperatures is best characterised by time-dependent correlation functions and transport coefficients. We have mostly focused on single-particle motion which is sampled with $N$ times better statistics than collective dynamics in the simulations. Examples of the normalised velocity autocorrelation function $Z(t) = \langle \vect{v}(t) \cdot \vect{v}(0) \rangle/\langle |\vect{v}|^2 \rangle$, where $\vect{v}(t)$ is the velocity of a given particle at time $t$, are shown in \fig{F:Z} for several temperatures. As expected, $Z(t^*)$ is seen to relax more and more rapidly as $T^*$ decreases. However, just below the percolation threshold at $T_p^*=0.177$, $Z(t^*)$ begins to exhibit a striking oscillatory behaviour.  Oscillations in $Z(t)$ normally involve periods of negative values, which signal the backscattering of particles in a dense fluid as the particles, on average, reverse their velocity.  In contrast, our low-density gel exhibits a $Z(t)$ that oscillates as it goes to zero without changing sign, except at the very lowest $T^*$.  Two frequencies can be discerned.  The shorter period (0.3 to $0.5t_0$) can be attributed to the one-dimensional rattling of particles within the chains.  This assignment is compatible with the Einstein frequency derived from a Taylor expansion to the quadratic term \cite{Dinsmore06a} of the potential energy of a dumbbell moving in one dimension between two fixed dumbbells with aligned dipoles.  This rapid oscillation is superimposed upon a slower movement of period $1.8t_0$ corresponding to concerted movement of a chain segment \cite{DelGado07a}.

\begin{figure}[tb]
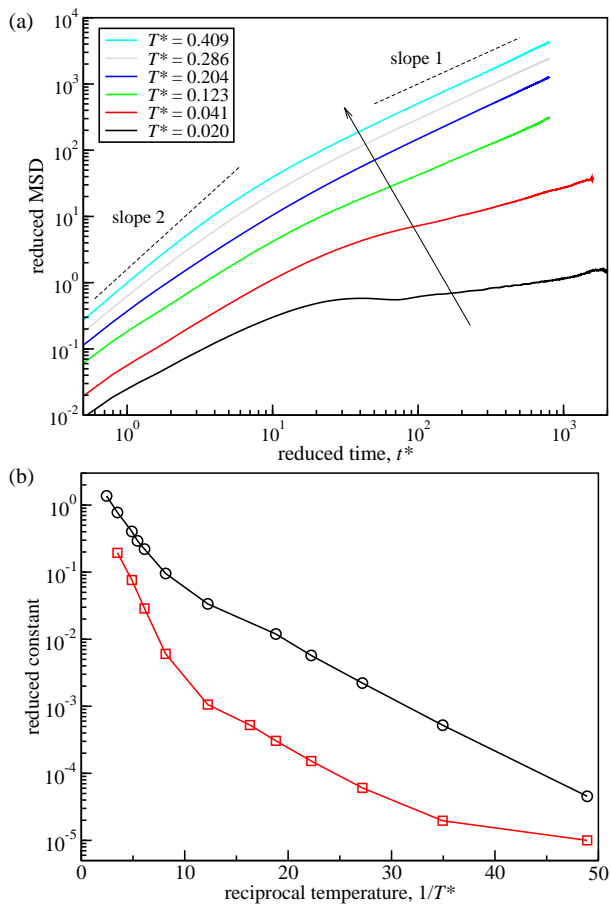

\centerline{\includegraphics[width=80mm]{fig4a}}
\centerline{\includegraphics[width=80mm]{fig4b}}
\caption{
(a) Reduced mean-square displacement (MSD) as a function of time.  Dashed lines indicate log--log slopes of $2$ (ballistic regime) and $1$ (diffusive).  The arrow intersects the curves in order of increasing temperature.
(b) Arrhenius plot of the reduced diffusion constant $D^*$ (circles) and reduced rate constant $k_b^*$ for bond breaking (squares).
}
\label{F:diff}
\end{figure}

Figure \ref{F:diff}(a) shows the mean-square displacement (MSD) of a particle from its original position, $\langle |\vect{r}(t) - \vect{r}(0)|^2\rangle$ versus time, on a double-logarithmic plot.  At the higher temperatures the MSD goes over directly from the initial ballistic regime ($\sim t^2$) to the linear Einstein regime at longer times \cite{DelGado07a}.  Below $T_p^*$ a sub-diffusive regime, characterised by a clear inflection point sets in at intermediate times.  The sub-diffusive regime extends over a rapidly increasing time interval as $T^*$ drops, and at the lowest $T^*$ the normal diffusive regime is only just reached within the time-scale of our simulation.  In other words, the system becomes dynamically arrested, at least over the accessible time scale, which covers several decades.  The particle diffusion constant $D^*$ is given by the slope of the MSD in the linear regime or by the time integral of $Z(t)$ \cite{Book:Hansen-McDonald3} and drops rapidly with $T^*$, as illustrated in \fig{F:diff}(b).  At low temperatures, where diffusion is an activated process, $D^*$ follows an Arrhenius law.

Another instructive diagnostic, also plotted in \fig{F:diff}(b), is provided by the rate of bond breaking.  Defining $b_{ij}(t)$ to be unity if dumbbells $i$ and $j$ are neighbours at time $t$ and zero otherwise, we may construct a bond correlation function $B(t)=\langle b(t)b(0)\rangle$, which exhibits a regime of exponential decay over a wide range of temperatures, allowing a reduced rate constant $k_b^*$ to be defined.  As seen for $D^*$, $k_b^*$ decreases rapidly upon cooling from high temperatures, and enters a regime of Arrhenius temperature dependence once the network is properly established.  However, at very low temperatures another change is seen, and $k_b^*$ is higher than an extrapolation would predict.  The relative ease of breaking bonds here can be understood by noting that a dipolar particle can only make two energetically optimal bonds (i.e., with dipoles aligned head-to-tail).  The increasing average coordination number at low temperature necessarily means that many particles are forming bonds with frustrated geometries, which require less energy to break.

\begin{figure}[tb]
\centerline{\includegraphics[width=80mm]{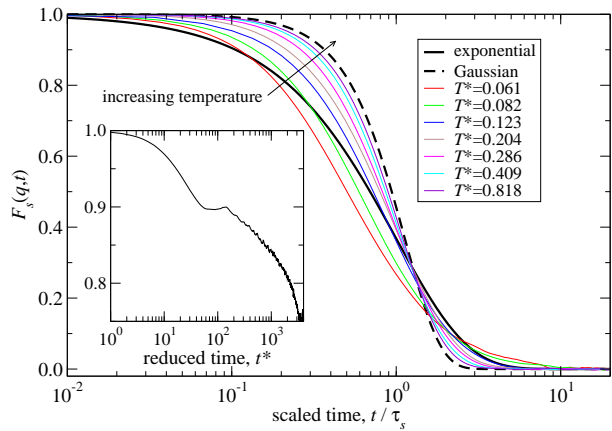}}
\caption{Self-intermediate scattering function $F_s(q,t)$ at $q\sigma = 1.12$ as a function of time scaled by the relaxation time $\tau_s$ at each temperature.  Inset: $F_s(q,t)$ at $T^*=0.020$ as a function of reduced time $t^*$.}
\label{F:Fs}
\end{figure}

The local motion of individual particles is characterised by the the self-intermediate scattering function of incoherent neutron scattering experiments \cite{Book:Hansen-McDonald3}
\begin{equation}
\label{E:Fs}
F_s(q,t) = \frac{1}{N} \sum_{j=1}^N \left \langle e^{{\rm i} \vect{q} \cdot [\vect{r}_j(t) - \vect{r}_j(0)]} \right \rangle
\end{equation}
where $\vect{q}$ is a wavevector compatible with the periodic boundary condition in the simulations. A $q$-dependent relaxation time $\tau_s(q)$ may be defined by the time-integral of (\ref{E:Fs}). At high temperatures, a ballistic regime is approached and $F_s(q,t)$ shows Gaussian decay \cite{Book:Hansen-McDonald3}.  This is illustrated in \fig{F:Fs}, where $F_s(q,t)$ is plotted versus $t/\tau_s(q)$ for $q\sigma = 1.12$.  For $T < T_p^*$, the correlation function deviates increasingly from the Gaussian limit, but never conforms to the exponential decay that would be expected in a dense liquid.  At sufficiently low temperature, a stretched long-time tail develops.  At the lowest temperature investigated, $T^*=0.020$ (inset), $F_s(q,t)$ exhibits strong dynamical slowing down, with an initial $\beta$ relaxation, followed by a plateau (which defines a ``non-ergodicity'' parameter $f_q \simeq 0.895$) and a final $\alpha$-relaxation for $t^*>170$, using the standard terminology of the kinetic glass transition \cite{Book:Hansen-McDonald3,Gotze:1992RPP}. Qualitatively similar behaviour is observed for the collective intermediate scattering function (or density autocorrelation function) $F(q,t)$, although the latter is subject to much larger statistical noise, and requires averages to be taken over several replicas, i.e., different initial conditions of the system.

\begin{figure}[tb]
\centerline{\includegraphics[width=80mm]{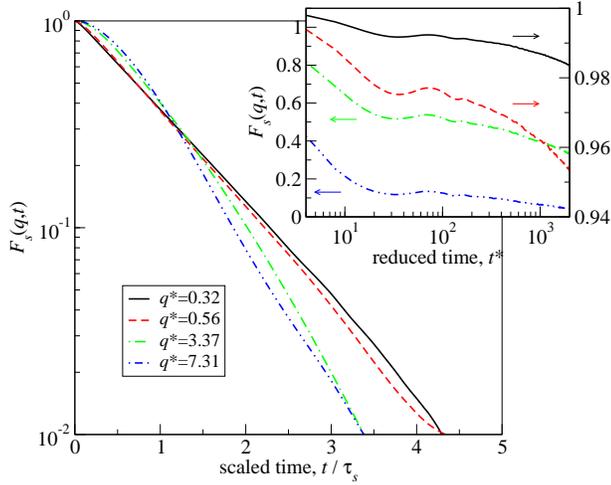}}
\caption{Dependence of the self-intermediate scattering function $F_s(q,t)$ on the wavevector $q$ at T=0.164 as a function of time scaled by the relaxation time $\tau_s$ at each wavevector.  Inset: $q$-dependence of $F_s(q,t)$ at $T^*=0.020$ as a function of reduced time $t^*$.  Arrows associate each curve with either the left-hand or the right-hand vertical scale.}
\label{F:qdep}
\end{figure}

The dependence of the self-intermediate scattering function on the wavevector $q$ is illustrated in \fig{F:qdep} at $T^*=0.164$ and $T^*=0.020$ (inset).  The former temperature lies just below $T^*_p$, where a regime of exponential decay is rapidly reached at all wavevectors, indicating that the preliminary formation of the network is not associated with any dynamical arrest.  The initial curvature of the relaxation at large $q$ is due to the very brief ballistic travel of particles before they reach their neighbour.  This region is emphasised for large $q$ because the time axis has been scaled by $\tau_s(q)$, which decreases rapidly with $q$.  In contrast, at the lower temperature (inset of \fig{F:qdep}), $F_s(q,t)$ encounters a $q$-dependent plateau, typical of glassy dynamics and long-lived dynamical arrest.

The data shown in the inset of \fig{F:qdep} are semi-quantitatively reproducible with different initial conditions, hinting that full dynamical arrest may only occur in the zero-temperature limit in low-density gels, contrary to the case of high-density structural glasses which exhibit clear-cut non-Arrhenius behaviour at finite temperatures.  This qualitative difference in behaviour may be attributed to the fact that particles are not completely caged by neighbours in the open network, and that ``floppy'' motions of the latter are not strongly inhibited at low packing fractions.

In summary our simulations predict that dilute dispersions of dumbbell-shaped, dipolar colloidal particles undergo a two-step gelation transition, with a reversible, ergodic network gel formed at a percolation threshold, characterised by relatively short-lived bonds, as in ``living'' polymers \cite{Cates:1987MAC}, followed by a kinetic transition to a ``frozen'', ergodicity-breaking gel at lower temperatures. The predicted dynamical behaviour is not unlike that observed in very different colloidal systems involving short-range attractive forces between particles \cite{Segre01a, Campbell05a, Manley05a, DelGado05a, Zaccarelli06a}, thus hinting at a certain universality of structural arrest for colloidal gels.

The model put forward in this letter could be mimicked experimentally using suspensions of sterically stabilised, oppositely charged colloidal particles assembled in pairs by the emulsion technique developed by Pine and collaborators \cite{Manoharan:2003SC,Zerrouki:2006LANG}.  The wide variation of $T^*$ could be achieved by tuning the particle dipole moment $\mu$, e.g., by varying the surface charge.  Future work will investigate the dependence of the present gelation mechanism on the colloidal volume fraction, which is a more convenient control parameter in experiments.

\acknowledgments
The authors thank Prof.~Francesco Sciortino and Prof.~Walter Kob for their insightful comments on the original manuscript.  R.B.~acknowledges the support of EPSRC within the Portfolio Grant number RG37352.  M.A.M.~is grateful to Churchill College, Cambridge and EPSRC for financial support.  He also thanks Prof.~Daan Frenkel and Dr.~Dima Lukatsky for stimulating his interest in dipolar systems.

\end{document}